\documentstyle[prl,aps,preprint,epsfig]{revtex}

\tightenlines

\begin{document}

\date{\today}
\draft

\title{
A \underline{k}-space transport analysis of the BEEM
spectroscopy of Au/Si Schottky barriers
}

\author{
U. Hohenester, P. Kocevar
}

\address{
Inst. f. Theoretische Physik, 
Karl-Franzens-Universit\"at Graz,
A-8010 Graz (AUSTRIA)
}

\author{
P.L. de Andres and F. Flores
}

\address{
Instituto de Ciencia de Materiales (CSIC), 
Cantoblanco, E-28049 Madrid (SPAIN) \\
Dept. de Fisica Teorica de la Materia Condensada (UAM), 
Universidad Autonoma de Madrid, E-28049 Madrid (SPAIN)
}

\maketitle

\begin{abstract}
We address the question of the spatial resolution of
ballistic electron emission microscopy (BEEM) of
Shottky barriers in Au(111)/Si(100) and Au(111)/Si(111)
interfaces. A novel combination of Green-function and
\underline{k}-space Ensemble-Monte-Carlo techniques is used to
obtain new insights into the spatial and energetic
evolution of the STM-tip-induced electrons during
their passage through the metallic layer before reaching
the metal-semiconductor interface. In particular,
it is shown how the effect of band-structure-induced
directional focusing of the electrons enforces a
reinterpretation of existing experimental data.
\end{abstract} 

REF.: Proc. 10th Conf. on Microscopy of Semiconducting Materials MSM-X
(Oxford April-1997), Ed. T. Cullis, in print.

\section{Introduction}

The recent theoretical prediction (de Andres et al. 1997)
of decisive band-structure effects in the propagation of
the STM-tip-induced hot electrons through the metal layer
brought a new facet to the discussion about the very high
spatial resolution of ballistic electron emission 
microscopy (BEEM) of Au/Si Shottky barriers
(Bell 1996; Prietsch 1995). While most interpretations
of BEEM data on Au/Si(100) and Au/Si(111) interfaces
have assumed a narrow forward cone of
tunneling-injected electrons and explained the very
similar spectra and their high spatial resolution
through various forms of collisional beam broadening,
the prediction of a pronounced off-axis shift and
broadening of the angular distribution just below
the surface reopens the discussion about the role
of the band structure and of scattering processes
in the bulk and at the boundaries of the metallic
layer. It is the purpose of the present analysis
to improve the conventional energy-space descriptions
and Monte-Carlo simulations of the hot-electron
dynamics by providing a detailed \underline{k}-space
Ensemble-Monte-Carlo simulation of the passage
of the hot electron through the Au layer,
including the essentials of the band structure
in the directional spectrum of the injected electrons,
in the free-particle propagation, and also in the
scattering cross sections for electron-electron
(e-e), electron-phonon (e-ph), and electron-boundary
scattering. The recent experimental data of Bell
(1996) for varying layer thickness d and temperature
will be reanalysed and the results contrasted with
Bell's original interpretation. It will be shown
that the two most intensively debated questions
about BEEM spectroscopies of the Au/Si system,
namely the questions about the origin of the
great similarity between Au/Si(100) and Au/Si(111)
spectra and about the extreme high spatial resolution,
can, for the first time, be directly answered
without the use of adjustable parameters or
ad hoc assumptions.

\section{Transport model}

Before presenting our transport model, we briefly
summarise the conventional model, which is
mainly based on the ideas of Kaiser and Bell (1988).
The original KB model assumes (i) that the injected
distribution at the metal surface is concentrated
within a narrow forward cone, (ii) that the
k-distribution at the metal/semiconductor interface
is identical to the injected k-distribution at
the metal surface, i.e. the $\underline{k}$-vector
parallel to the plane 
$\underline{k}^{\parallel} \approx 0$,
and (iii) that
$\underline{k}^{\parallel}$ 
is conserved at
the interface (i.e. specular reflection/transmission
via continuity of wavefunctions). As a consequence
of the different orientations of the six
conduction-band valleys in Si with respect to
the impinging narrow forward cone at the interface,
BEEM spectra for Au/Si(100) and Au/Si(111) should,
in the absence of strong scattering effects,
be distinctly different. This should occur
because of matching of 
$\underline{k}^{\parallel}$ 
for Au/Si(111), a prediction in strong contrast
with the experimental facts.

We now turn to our present transport model. We first
note that STM and LEED studies show that Au films
grow on Si(100) and Si(111) by forming crystals
oriented preferentially in the $\lbrack 111 \rbrack$
direction. Then the empirical-tight-binding
Green function analysis (Garcia-Vidal et al. 1996)
of the coherent electron propagation from the
STM tip through the tunneling gap into the
metal layer reveals that the STM electrons
achieve their bulk Bloch character, with propagation
gaps due to forbidden regions of phase space,
after passing roughly $20$ {\AA} within the
metal. The injected distribution at
$z=20$ {\AA} turns out to reach its maximum
at the edge of the planar Brillouin zone at
$\approx 30$ degrees (from the normal direction),
with an average $1/ \cos \theta$ distribution
law (de Andres et al. 1977). We should stress
that the detailed shape of this distribution
depends on the exact tip-surface configuration.
{\bf This angular distribution is essentially
different from the conventionally assumed
narrow forward cone and should drastically
change most of the previous interpretations
of BEEM data on Au/Si}.
The energetic spectrum of the injected electrons
is taken from conventional planar tunneling 
theory (Prietsch 1995).

In view of the fact that the total mean free path
is much greater than $20$ {\AA}, our Monte-Carlo
simulations of the electronic scattering
dynamics use the above Green function result as
the input ensemble of injected STM electrons at
the surface. Appropriately modifying
well-established Ensemble-Monte-Carlo techniques
for the solution of the non-linear steady-state
Boltzmann equation for semiconductors
(Hohenester et al. 1992), the hot-electron
distribution function
$f_{IF}(\underline{k})$ at the interface is
obtained as follows. Starting from quasifree
electrons ($m_{eff}=m_{0}$), we correct for
band-structure effects on the electron propagation
by cutting off the forbidden directions arising
from gaps in the constant-energy surfaces. For
our case of injection energies about $1$ eV
above the Fermi energy, these "propagation gaps"
form cones with an opening angle of $10$ degrees
around the $\lbrack 111 \rbrack$ directions and are easily
included in the scattering dynamics by use of
Monte-Carlo rejection techniques.

The total and differential cross sections for the
scattering between the hot electrons and those
of the "cold" metallic background are treated
via a dynamically screened Coulomb potential
(Pines 1968), via the standard Monte-Carlo
procedure (Jacoboni and Reggiani 1983).
This full \underline{k}-space description
should be contrasted with the earlier MC
simulations of the bulk scattering dynamics
in the metal (Bauer et al. 1993, Bell 1996),
which are based on an energy-space description,
with mean free paths numerically adjusted to the
experimental data by use of simple rational
functions of energy.

Assuming specular transmission/reflection
(via wavefunction matching at a step-like
barrier $\Theta_{B}$) and either specular
or diffuse reflection at the free metal surface
(both types of reflections resulting in
practically identical simulated BEEM currents),
the boundary scatterings are treated in the
conventional way (Bauer et al. 1993, Bell 1996).

The simulation of each electron is also followed
in \underline{r}-space and stopped after it has
passed the interface (i.e. when $z>d$) or when
its energy has dropped below the top of the
barrier. In this way one obtains the energetic
and angular distribution of transmitted electrons
at the interface. We further assume negligible
current modifications within the semiconductor,
which should be well justified for the modest
electron energies of our present concern
(Prietsch 1995). Then the relative portion
of transmitted electrons directly determines
the relative BEEM current $I_{B}/I_{T}$
as function of the tunnel bias $V_{T}$
for the given barrier $\Theta_{B}$. Four
our calculations of the $I_{B}$ characteristics
and analysis of the spatial resolution we used
the standard value $\Theta_{B}=0.8$ eV; we
checked that the known small temperature
variation of $\Theta_{B}$ does not change
the essentials of our results.

\section{Results and conclusions}

Although we have analysed both Si orientations, we
concentrate our following discussion mainly on
Au(111)/Si(111). We first state that, as had
already been demonstrated by GF calculations of
Garcia-Vidal et al. (1996) and de Andres et al.
(1997) for a pure ballistic electron propagation
through the metal layer, the comparable
$I_{B}$ thresholds and magnitudes for
Au/Si(111) and Au/Si(100) and the high spatial
BEEM resolution are a direct consequence
of the band structure-induced non-forward
electron injection. This claim can now be
substantiated by the results of our inclusion
of the scattering dynamics in the former
free-electron scenario of de Andres et al. (1997).
To demonstrate clearly the effect of scattering
processes, we first consider the very narrow
off-axis initial distribution originally
obtained by Garcia-Vidal et al. (1996)
within the present GF approach by neglect
of self-interference effects in the coherent
free-electron propagation. Figure 1 shows the
resulting lateral current distribution within
the layer (taken as semi-infinite) at three
typical penetration depths (left) and its
much smaller fraction due to scattered
electrons (right; note change of scale).
One can easily distinguish the build up
of secondary "hot" electrons at the
lowest energies due to inelastic e-e
scatterings and the angular spreading
of the distribution through the quasi-elastic
e-ph interactions. For this illustrative
example a high spatial resolution in
$I_{B}$ would be found, caused by the
dominance of those "happy" \underline{k}-matching
electrons (distributed typically between
35 and 45 degrees within the high-angle
wing of the distribution) which cross
the interface at their first attempt.

Turning to the initial distribution underlying this study
(de Andres et al. 1997), our simulations also
yield a high spatial resolution, which again is
caused by the fraction of "happy" electrons
in the angular range between 35 and 45 degrees,
which now lies in the low-angle wing of the
distribution. Figure 2 compares our results for
the BEEM current characteristics (full lines)
with Bell's experimental data (diamonds)
for two different layer thicknesses and
temperatures. The theoretical curves are 
in quantitative agreement with the data,
except for the case of "thick" layers
($d=300$ {\AA}) and "low" temperatures
($77$ K) (i.e. for thickness on the order of
the mean free path for inelastic and
quasielastic scatterings and for a temperature
with strongly reduced e-ph scatterings).
At present we have no explanation for this
pronounced discrepancy. We can only suspect
that some dynamical details, in particular
regarding the interface dynamics, are still
missing in the present simulation scenario
and become decisive in thick layers and at
low temperatures. We should note that Bell (1996)
has attempted to explain his experimental
finding of a decreasing low-temperature BEEM
current with increasing layer thickness by the
decrease of multiple internal reflections and
the corresponding decrease of the number of
"attempts for transmission" at the interface.
Our simulations confirm this dominance of
multiple reflections in thin layers, but
cannot reproduce the decrease of $I_{B}$
with increasing d, because our theory
lacks the $\underline{k}^{\parallel}$-matching
restrictions of Bell's forward injection
scenario. Moreover, our reproduction of
Bell's simulations revealed that his
interpretation and the practically perfect
agreement of his theoretical $I_{B}/I_{T}$
versus $V_{T}$ characteristics
(dashed lines in Fig. 2) with the experimental
data strongly depend on his choice of mean
free paths and of the detailed injected
current distribution: the calculated $I_{B}$
changes drastically, (i) if the opening
angle is e.g. changed from $10$ to $20$ degrees,
or (ii) if the energetic window is changed
from Bell's $0.2$ to $0.4$ eV, or
(iii) if his mean-free path description
is replaced by our detailed bulk-scattering
dynamics in \underline{k}-space. So we
believe that no convincing explanation
exits for the decrease of $I_{B}$
with increasing d at low temperature.

To summarize, the present \underline{k}-space
description of the injection spectrum and of
the bulk scattering dynamics has no adjustable
parameter and therefore improves over the
many-parameter fits of earlier theoretical
interpretations. It turns out that scattering
processes have no decisive influence on the
spatial BEEM resolution. Although quantitative
agreement with experiment is found in most cases,
some remaining discrepancies seem to indicate
the need for an improved description of the
electron dynamics at the interface.

\begin{figure}
\epsfig{figure=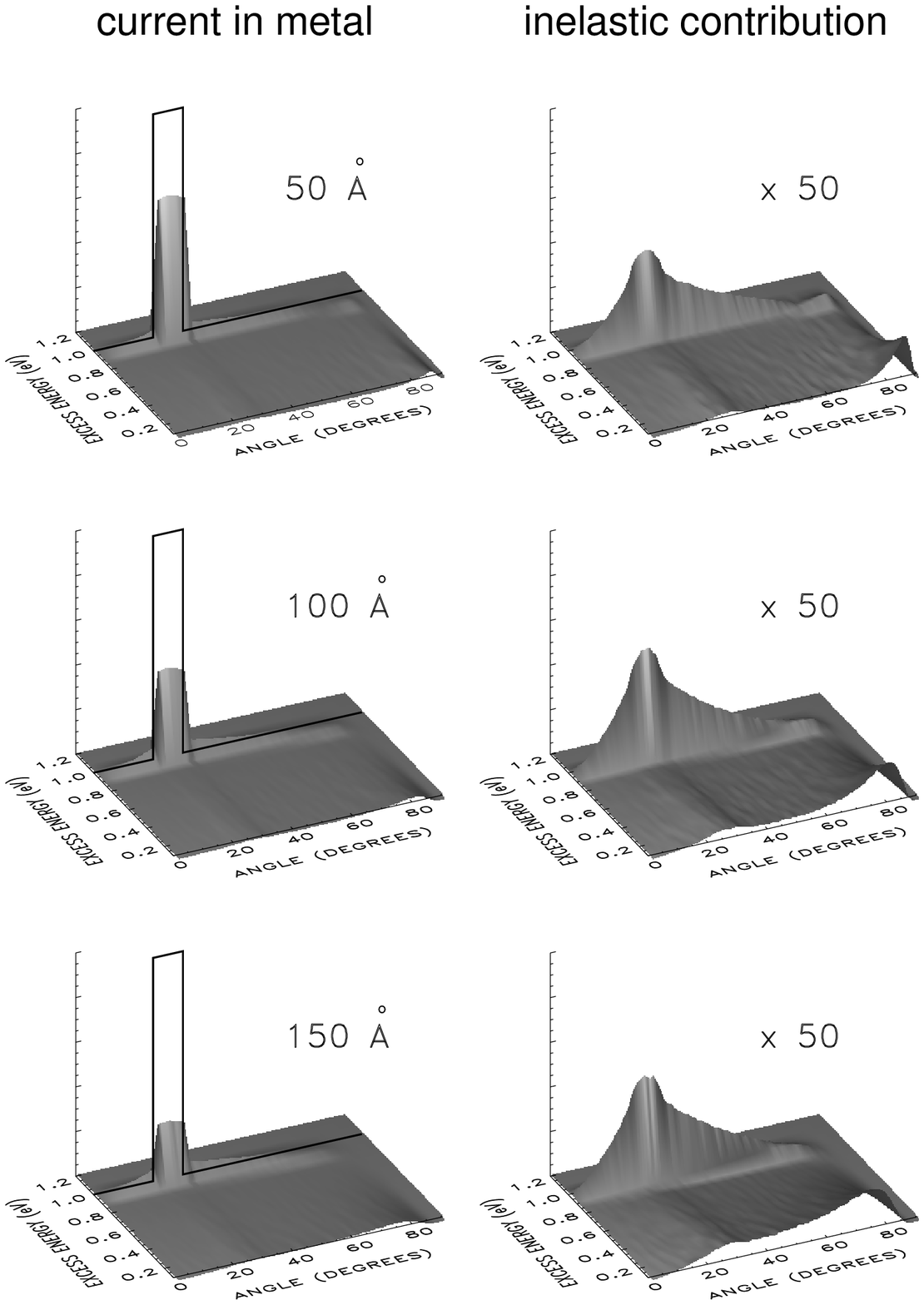,height=21cm,width=15cm}
\caption{
Spatial evolution of current density in metal layer
for injected distribution (at z=0) shown as solid line.
}
\end{figure}

\begin{figure}
\epsfig{figure=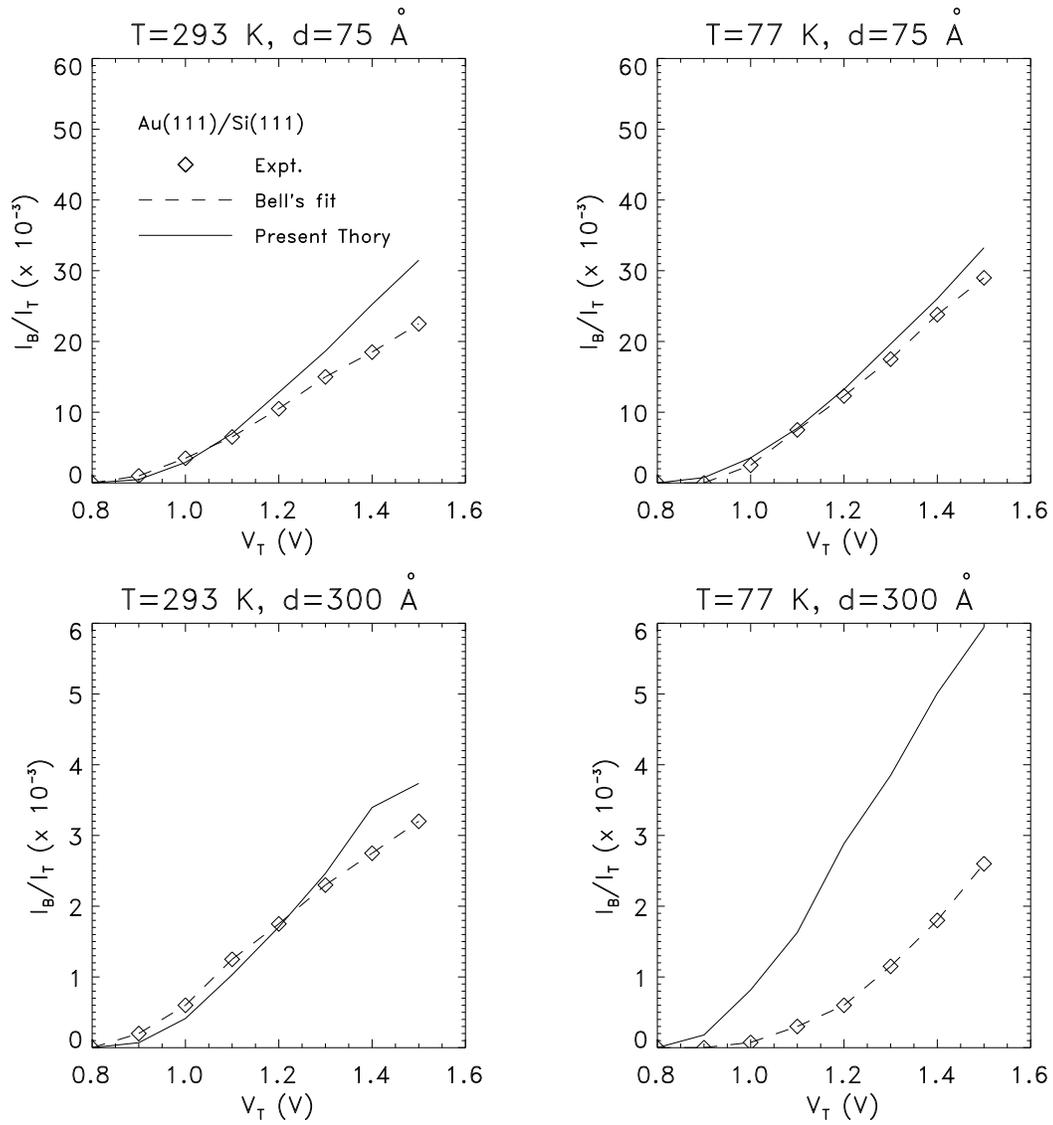,height=15cm,width=15cm}
\caption{
Relative BEEM current versus bias voltage;
experimental data taken from Bell 1996
}
\end{figure}

\end{document}